\documentclass{ifacconf}

\usepackage{graphicx}      
\usepackage{natbib}        
\usepackage{amsmath}      
\usepackage{mathrsfs}

\usepackage{amssymb} 

\newtheorem{lemma}{Lemma} 
\newtheorem{theorem}{Theorem}
\newtheorem{definition}{Definition}
\usepackage{algorithm}
\newtheorem{assumption}{Assumption}
\newtheorem{remark}{Remark}

\usepackage{booktabs}
\usepackage{algorithmic}

\newcommand{\pt}{\phi_k^{\mathrm{T}}\theta_k} 
\newcommand{\ptb}{\phi_k^{\mathrm{T}}\bar{\theta}_k}
\begin{document}
\begin{frontmatter}


\title{$\ell_1$-Based Adaptive Identification under Quantized Observations with Applications\thanksref{footnoteinfo}} 
\thanks[footnoteinfo]{This paper was supported by the National Key Research and Development Program under Grant No. 2024YFC3307200 and the National Natural Science Foundation of China under Grant No. 12288201}

\author[1,3]{Xin Zheng} 
\author[2,1,3]{Yifei Jin} 
\author[1]{Yujing Liu}
\author[1,3]{Lei Guo}

\address[1]{State Key Laboratory of Mathematical Sciences, Academy of Mathematics and Systems Science, 
Chinese Academy of Sciences, Beijing 100190, China. 
(e-mail: \{zhengxin2021, liuyujing, lguo\}@amss.ac.cn).}

\address[2]{School of Advanced Interdisciplinary Sciences, University of Chinese Academy of Sciences, Beijing 101408, China. (e-mail: jinyifei@amss.ac.cn).}

\address[3]{School of Mathematical Sciences, University of Chinese Academy of Sciences, Beijing 100049, China.}

\begin{abstract}                
Quantized observations are ubiquitous in a wide range of applications across engineering and the social sciences, and algorithms based on the $\ell_1$-norm are well recognized for their robustness to outliers compared with their $\ell_2$-based counterparts. Nevertheless, adaptive identification methods that integrate quantized observations with $\ell_1$-optimization remain largely underexplored. Motivated by this gap, we develop a novel $\ell_1$-based adaptive identification algorithm specifically designed for quantized observations. Without relying on the traditional persistent excitation condition, we establish global convergence of the parameter estimates to their true values and show that the average regret asymptotically vanishes as the data size increases. Finally, we apply our new identification algorithm  to a judicial sentencing problem using real-world data, which demonstrates its superior performance and practical significance.
\end{abstract}

\begin{keyword}
Adaptive identification, Quantized observations, $\ell_1$-norm optimization, Global convergence.
\end{keyword}

\end{frontmatter}

\section{Introduction}
\label{submission}

Quantized observations are prevalent across numerous critical domains (\cite{wang2024asymptotically}), including judicial sentencing (\cite{Bai2021Doctrinal, Peng2021DistributiveJustice}), medical systems (\cite{caodecentralized, li2025federated}), financial risk assessment (\cite{torres2020classification, OET20134510, wang2022imbalanced}), and autonomous driving (\cite{dargie2010fundamentals, 9827020, katare2024analyzing}). For instance, due to privacy constraints, data collected from different medical institutions are often quantized and exhibit non-independent and non-identically distributed (non-i.i.d.) characteristics (\cite{caodecentralized}), while in wireless sensor networks, each low-cost sensor can only provide quantized measurements with limited bits because of power and bandwidth constraints (\cite{dargie2010fundamentals}). These examples illustrate the ubiquity and practical importance of quantized data in both engineering and social science applications.

It is well recognized that learning algorithms based on the $\ell_1$-loss function are generally more robust to outliers than those based on the widely used $\ell_2$-norm (\cite{ghosh2017robust}). 
Motivated by this robustness and the ubiquity of quantized observations in practical systems, designing $\ell_1$-norm-based adaptive identification algorithms for quantized data is of particular importance. 
However, the inherent nonlinearity of quantized observations, together with the nondifferentiability of the $\ell_1$-norm, makes the theoretical analysis of such algorithms particularly challenging. 
Consequently, research that integrates quantized observations with $\ell_1$-optimization remains limited. 
Constructing an $\ell_1$-based adaptive identification framework for quantized observations is thus of both theoretical and practical value.

Recent studies have made progress on $\ell_2$-norm-based adaptive identification for quantized systems under relaxed excitation data conditions. For example, \cite{dai2024adaptive} developed an adaptive identification algorithm that achieves convergence without relying on the persistent excitation (PE) condition, and \cite{KE2024105941} addressed the joint identification of unknown system and noise parameters in quantized systems under non-PE conditions, establishing corresponding identifiability results. Inspired by these works, we aim to develop a more robust $\ell_1$-norm-based adaptive identification algorithm that can achieve reliable convergence without PE conditions.

To this end, we propose a new $\ell_1$-based adaptive identification algorithm specifically designed for quantized observations. The proposed algorithm integrates an $\ell_1$-loss function with a two-step quasi-Newton scheme to enhance robustness and convergence. Without requiring traditional PE conditions, we prove that the parameter estimates globally converge to the true values and that the average regret asymptotically vanishes as the data size increases. Finally, an experiment on a real-world judicial sentencing dataset demonstrates that our algorithm outperforms commonly used $\ell_2$-based methods, highlighting its superior performance and practical significance.

The main contributions of this paper are summarized as follows:
\begin{itemize}
\item We develop a new $\ell_1$-norm-based adaptive identification algorithm for quantized observations, combining an $\ell_1$-loss with a quasi-Newton updating scheme.
\item We establish theoretical guarantees under non-PE conditions, proving global convergence of the parameter estimates and vanishing average regret, which broadens applicability to feedback systems.

\item We demonstrate the effectiveness of the proposed algorithm through experiments on a real-world sentencing dataset, showing superior predictive performance compared with $\ell_2$-based counterparts.
\end{itemize}

The remainder of this paper is organized as follows. Section~\ref{problem formulation} presents the quantized observation model, notations, and assumptions. Section~\ref{main results} introduces the proposed algorithm and main theoretical results. Section~\ref{Experiment} demonstrates the experimental validation, and Section~\ref{conclusion} concludes the paper.

\section{Problem Formulation} \label{problem formulation}

In this section, we will first give the model, followed by the notations and assumptions.

 \subsection{Quantized Observation Model} \label{SCM}

We consider the following general stochastic quantized observation model:
\begin{gather}\label{sclassmodel}
	y_{k+1} = S_k(\phi_k^{\mathrm{T}}\theta+\epsilon_{k+1}), \quad k = 0, 1, 2, \cdots,
\end{gather}
\noindent where \(y_{k+1} \in \{1, 2, 3, \cdots, m+1\}\) represents the observation, \(m+1\) denotes the total number of quantized values, \(\phi_k \in \mathbb{R}^d\) (with \(d \geq 1\)) is the stochastic regression vector, and \(\epsilon_{k+1} \in \mathbb{R}\) is a random noise, \(\theta \in \mathbb{R}^d\) is an unknown parameter vector to be estimated, \(S_k(\cdot): \mathbb{R} \rightarrow \mathbb{N}\) is a non-decreasing saturation function defined as follows:
\begin{gather} \label{sfunction}
	S_k(x)=\left\{\begin{array}{cl}
	1, & x \leq c_{1k}, \\
        \cdots & \cdots \\
	m, & c_{(m-1)k} < x \leq c_{mk},\\
        m+1, & x > c_{mk}, 
	\end{array}\right.
\end{gather}
where \(\{c_{ik}, ~i=1,\cdots, m\}\) denotes the set of thresholds.

\begin{remark}
The model \eqref{sclassmodel} provides a general framework applicable to various quantized systems (\cite{wang2024asymptotically, dai2024adaptive, KE2024105941}). Moreover, any strictly increasing sequence of quantized values can be employed in practice, while the range from $1$ to $m+1$ is adopted here for simplicity.  Notably, the model reduces to the commonly encountered binary-valued observations case when setting $m=1$.
\end{remark}

In this paper, we adopt the following $\ell_1$-norm loss to measure the accuracy of prediction: 
\begin{gather}\label{classloss}
	\frac{1}{k}\sum\limits_{i= 1}^{k} \vert y_{i+1} - \hat{y}_{i+1}\vert,
\end{gather}
where  $k$ denotes the number of online samples arrived, \(\hat{y}_{i+1}\) is an adaptive classifier that tries to minimize the loss \eqref{classloss}. 

\subsection{Notations and Assumptions} \label{Notations and Assumptions}
The following notations and assumptions will be needed.

\noindent\textbf{Notations.}  \(\|\cdot\|\) denotes the the Euclidean norm of a matrix or vector; $|\cdot|$ denotes the $\ell_1$ norm of a vector which is the summation of the absolute values of its components; For a matrix \(M\), \(\lambda_{\min }\{M\}\) and \(\lambda_{\max}\{M\} \) denote the minimum  and maximum eigenvalues of \(M\) respectively, \(M^{\mathrm{T}}\) denotes the transpose of \(M\); \(\{\mathcal{F}_k, k\geq 0\}\) is a sequence of non-decreasing \(\sigma\)-algebras, and \(\mathbb{E}[\cdot \mid \mathcal{F}_k]\) denotes the conditional expectation; For simplicity, the notations \(\mathbb{E}_k[\cdot]\) or \(\mathbb{E}_k\{\cdot\}\) may be used in place of \(\mathbb{E}[\cdot \mid \mathcal{F}_k]\); \(\{x_k, \mathcal{F}_k, k\geq 0\}\) is called an adapted sequence if the random variable \(x_k\) is \(\mathcal{F}_k\)-measurable for all \(k \geq 0\); For two real sequences $\{a_k, k\geq0\}, \{b_k, k \geq 0\}$ with $b_k>0$, $a_k =o(b_k)$ means that $a_k/b_k \rightarrow 0$ as $k \rightarrow \infty$,  $a_k=O(b_k)$ means that there is a positive constant $C$ such that $|a_k| \leq C b_k$ for all $k>0$; \(\operatorname{sgn}[\cdot]\) is  the sign function and \(\operatorname{I}_{[\cdot]}\) is the indicator function.

 Moreover, we introduce the following assumptions:
 \begin{assumption} \label{threshold}
The thresholds \(\{c_{ik}, \mathcal{F}_k, k \geq 0\}_{ 1\leq i \leq m }\) are known adapted sequences, where the thresholds are non-decreasing in \(i\) and bounded in \(k\). In addition, there is a positive constant \(\Lambda_c\) such that:
\begin{gather} \label{upperlower}
    \sup_{k \geq 0} \max\{c_{mk}, -c_{1k}\} \leq \Lambda_c < \infty, ~\text{a.s.}
\end{gather}
\end{assumption} 

This assumption is relatively mild and serves to guarantee the identifiability of the parameters.

\begin{assumption}\label{A1}
The adapted sequence \( \{\phi_k, \mathcal{F}_k, k \geq 0\}\) is bounded, i.e., \(\sup\limits_{k \geq 0} \|\phi_k\| < \infty\), and the parameter vector \(\theta\) is an interior point of a known convex compact set \(\mathcal{K} \subseteq \mathbb{R}^d\).
\end{assumption}

By Assumption \ref{A1}, it is clear that there exists a bounded adapted sequence \(\{G_k, \mathcal{F}_k, k\geq 0\}\) such that
\begin{gather} \label{boundC}
   \sup_{x \in \mathcal{K}} |\phi_k^{\mathrm{T}}x| \leq G_k.
\end{gather}

\begin{assumption}\label{symmetry}
The conditional distribution function of the noise \(\epsilon_{k+1}\) given  \(\mathcal{F}_k\), denoted by \(F_{k+1}(\cdot)\), is known and satisfies \(F_{k+1}(0) = \frac{1}{2}\). Moreover, the corresponding conditional density function \(f_{k+1}(\cdot)\) is continuous. In addition, there exists a constant \(G\) such that \(G \geq \sup\limits_{k\geq 0} G_k\) and 
\begin{gather}\label{boundednessf}
\begin{aligned}
    0 &< \inf_{|x| \leq G+ \Lambda_c, ~k \geq 0} f_{k+1}(x) \\
    &\leq \sup_{|x| \leq G+ \Lambda_c,~k \geq 0} f_{k+1}(x) < \infty, \quad \text{a.s.}
\end{aligned}
\end{gather}
\end{assumption}

\begin{remark} \label{noise density requiered}
The conditional distribution and density functions of the noise \(\epsilon_{k+1}\) are essential for constructing the adaptive classification algorithm. This requirement is not restrictive, as the distribution can be estimated from real data using methods such as the noise distribution fitting (\cite{wang2022applications}). Moreover, the condition \( F_{k+1}(0) = \frac{1}{2} \) is imposed just for convenience in defining the optimal predictor in \eqref{classloss}. If instead \( F_{k+1}(a) = \frac{1}{2} \) for some \( a \neq 0 \), the optimal predictor can be adjusted to account for this shift without affecting our theoretical analysis. 
\end{remark}

Under Assumptions \ref{threshold}--\ref{symmetry}, by using Lemma 1 in \cite{zheng2024l1}, one can deduce that the best predictor of (\ref{classloss}) at time \(k\) is
\begin{equation} \label{predictc}
	\hat{y}_{k+1}^{*} =  S_k(\phi_k^{\mathrm{T}}\theta).
\end{equation}

Since the true parameter \(\theta\) is not known \textit{a priori}, it is necessary to develop an adaptive identification algorithm to estimate it. Once the estimate, denoted by \(\theta_k\), is obtained at time \(k\), it replaces \(\theta\) in \eqref{predictc}, giving the following adaptive predictor \eqref{(8)}:
\begin{gather}\label{(8)}
\hat{y}_{k+1} = S_k(\phi_k^{\mathrm{T}}\theta_k). 
\end{gather}

\section{The Main Results} \label{main results}
In this section, we first present the new algorithm based on the model \eqref{sclassmodel}, and then introduce the main theorems.

\subsection{Adaptive Algorithm}
To construct the adaptive algorithm, we need the following projection  operator \(\Pi_A(\cdot)\) :
\begin{definition}(\cite{zhang2022identification})\label{prodef}
The projection operator \(\Pi_A(\cdot)\) is defined as
\begin{equation}
	\Pi_A(x)=\underset{y \in \mathcal{K}}{\arg\min }\|x-y\|_A, \quad \forall x \in \mathbb{R}^d ,
\end{equation}
where \(\mathcal{K}\) is defined in Assumption \ref{A1}, \(\|z\|_A^2=z^T A z\) for any \(z \in \mathbb{R}^d\) and any positive definite matrix \(A\).
\end{definition}
This projection operator will ensure that the parameter estimates remain bounded during the computational process of the algorithm.

We are now ready to introduce the new $\ell_1$-norm adaptive identification algorithm. The details of this algorithm are outlined in  Algorithm \ref{algorithm1} below.

\begin{algorithm}
\caption{The Absolute Deviation Based Adaptive (ADA) Algorithm }
\label{algorithm1}
\textbf{Step 1.} Recursively compute the preliminary estimates \(\bar{\theta}_{k+1}\) for \(k \geq 0\):
\[
\begin{aligned}
\bar{\theta}_{k+1} =& \Pi_{\bar{P}_{k+1}^{-1}} \left\{ \bar{\theta}_k + \bar{a}_k \bar{P}_k \phi_k \bar{v}_{k+1} \right\}, \\
\bar{v}_{k+1} =& \operatorname{sgn}\left[y_{k+1} - S_k\left(\phi_k^{\mathrm{T}} \bar{\theta}_{k}\right)\right]  \\
&+\begin{cases}
       {\scriptstyle F_{k+1}\left(\bar{\delta}_{1k}\right) -1, \qquad \qquad \qquad \quad ~\text{if}~ \phi_k^{\mathrm{T}} \bar{\theta}_{k} \leq c_{1k}},  \\
     {\scriptstyle F_{k+1}\left(\bar{\delta}_{(i-1)k}\right)- \left[1 - F_{k+1}\left(\bar{\delta}_{ik}\right)\right]  
   , \quad\text{if } c_{(i-1)k} < \phi_k^{\mathrm{T}} \bar{\theta}_{k} \leq c_{ik}}, \\ 
    {\scriptstyle F_{k+1}\left(\bar{\delta}_{mk}\right), 
      \qquad \qquad \qquad \quad  ~ ~~\text{if }~ \phi_k^{\mathrm{T}} \bar{\theta}_{k} > c_{mk}},
\end{cases} \\
\bar{P}_{k+1} =& \bar{P}_k - \bar{a}_k \bar{\beta}_k \bar{P}_k \phi_k \phi_k^{\mathrm{T}} \bar{P}_k, \\
\bar{a}_k =& \frac{1}{\bar{\mu}_k + \bar{\beta}_k \phi_k^{\mathrm{T}} \bar{P}_k \phi_k}, \\
\bar{\beta}_k =& \inf_{\parallel x \parallel \leq \max \{ G_k + c_{mk}, G_k - c_{1k} \}} f_{k+1}(x),
\end{aligned}
\]
where $\bar{\delta}_{ik}  = c_{ik} - \phi_k^{\mathrm{T}} \bar{\theta}_{k}$, \(i=1, \cdots, m\), \(\left\{\bar{\mu}_k, \mathcal{F}_k, k \geq 0\right\}\) is a adapted sequence satisfying \( 0 < \inf\limits_{k \geq 0}\bar{\mu}_k \leq\sup\limits_{k \geq 0} \bar{\mu}_k<\infty \). The initial values \(\bar{\theta}_0 \) can be chosen arbitrarily from  \(\mathcal{K}\)  and with \(\bar{P}_0>0\).

\noindent
\textbf{Step 2.} Recursively compute the accelerated estimates \(\theta_{k+1}\)  given $\bar{\theta}_k$ at any time $k$:
\[
\begin{aligned}
\theta_{k+1} =& \Pi_{P_{k+1}^{-1}} \left\{\theta_k + a_k P_k \phi_k v_{k+1}\right\}, \\
v_{k+1} =& \operatorname{sgn}\left[y_{k+1} - S_k\left(\phi_k^{\mathrm{T}} \theta_{k}\right)\right] \\
&+\begin{cases}
   { \scriptstyle F_{k+1}\left(\delta_{1k}\right) -1, \qquad \qquad \qquad \quad ~\text{if}~ \phi_k^{\mathrm{T}} \theta_{k} \leq c_{1k}}, \\
    {\scriptstyle F_{k+1}\left(\delta_{(i-1)k}\right)- \left[1 - F_{k+1}\left(\delta_{ik}\right)\right]   
    }, \quad  {\scriptstyle\text{if } c_{(i-1)k} < \phi_k^{\mathrm{T}} \theta_{k} \leq c_{ik}}, \\ 
    {\scriptstyle F_{k+1}\left(\delta_{mk}\right), 
      \qquad \qquad \qquad \quad  ~ ~~\text{if }~ \phi_k^{\mathrm{T}} \theta_{k} > c_{mk}},
\end{cases} \\
P_{k+1} =& P_k - a_k \beta_k P_k \phi_k \phi_k^{\mathrm{T}} P_k, \\
a_k =& \frac{1}{\mu_k + \beta_k \phi_k^{\mathrm{T}} P_k \phi_k}, \\
\beta_k =& \scalebox{0.91}{$ 
\begin{cases} 
{\scriptstyle \frac{F_{k+1}\left(\delta_{1k}\right) - 
    F_{k+1}\left(\bar{\delta}_{1k}\right)}{\Delta_k} 
    \mathrm{I}_{[\Delta_k \neq 0]} + f_{k+1}\left(\delta_{1k}\right) 
    \mathrm{I}_{[\Delta_k = 0]},  ~~~~\text{if}~ \phi_k^{\mathrm{T}} \theta_{k} \leq c_{1k},} 
     \\[6pt]
    {\scriptstyle \left[\frac{F_{k+1}\left(\delta_{(i-1)k}\right) - 
    F_{k+1}\left(\bar{\delta}_{(i-1)k}\right)}{\Delta_k} + 
    \frac{F_{k+1}\left(\delta_{ik}\right)-F_{k+1}\left(\bar{\delta}_{ik}\right)}{\Delta_k} \right]
    \operatorname{I}_{[\Delta_k \neq 0]} } \\
     +{\scriptstyle [f_{k+1}(\delta_{(i-1)k}) + f_{k+1}(\delta_{ik})]  \operatorname{I}_{[\Delta_k = 0]}, \quad~~~~ \text{if } ~c_{(i-1)k} < \phi_k^{\mathrm{T}} \theta_{k} \leq c_{ik},}
     \\[6pt]
   {\scriptstyle \frac{F_{k+1}\left(\delta_{mk}\right) - 
    F_{k+1}\left(\bar{\delta}_{mk}\right)}{\Delta_k} 
    \mathrm{I}_{[\Delta_k \neq 0]} + 
    f_{k+1}(\delta_{mk}) \operatorname{I}_{[\Delta_k = 0]}, ~\text{if}~ \phi_k^{\mathrm{T}} \theta_{k} > c_{mk},} 
\end{cases} $}
\end{aligned}
\]
where $\delta_{ik}  = c_{ik} - \phi_k^{\mathrm{T}} \theta_{k}$, \(i=1, \cdots, m\), \(\Delta_k=\phi_k^{\mathrm{T}}\left(\bar{\theta}_k-\theta_k\right)\). $\left\{\mu_k\right\}_{k \geq 0}$ is defined analogously to $\left\{\bar{\mu}_k\right\}_{k \geq 0}$. The initial values \(\theta_0\) can be chosen arbitrarily from  \(\mathcal{K}\)  and with  \(P_0> 0 \).
\end{algorithm}

\begin{remark}
Algorithm \ref{algorithm1} is a two-step procedure, where the primary difference between the two steps lies in the convergence rate of parameter estimation: the second step converges faster than the first. This arises from the fact that we choose the adaptation gain \(\bar{\beta}_k\) as the infimum of the density function defined in Algorithm \ref{algorithm1}, ensuring the convergence of parameter estimation in the algorithm's first step. However, this choice of adaptation gain \(\bar{\beta}_k\) may cause the parameter estimation to converge slowly, as it could be very small, leading to extremely slow updates of \(\bar{P}_k\) and \(\bar{a}_k\). To address this issue, we introduce the second step of the algorithm to accelerate the convergence of the parameter estimation. In this step, a relatively larger adaptive gain \(\beta_k\) is chosen based on the parameter estimate \(\bar{\theta}_k\) obtained from the first step, enabling faster convergence rate. The impact of hyperparameter selection, follows the discussion in Remark 5 of \cite{zheng2024l1}. Therefore, we omit the details here. Although this two-step approach is inspired by \cite{zheng2024l1}, there are notable differences. The models in \cite{zheng2024l1} are primarily designed for predicting real-valued outcomes, whereas the current models are intended for a finite set of quantization levels, which leads to the key differences in the adaptive algorithms employed.
\end{remark}

\subsection{The Main Theorems} \label{main theorems}
Now we establish the asymptotic upper bounds for both the parameter estimation errors and the averaged regrets of the Algorithm \ref{algorithm1} under
general data conditions. 

We begin by establishing the global convergence result for the parameter estimates under non-PE conditions.

\begin{theorem}\label{mainThm}
If Assumptions \ref{threshold}--\ref{symmetry} hold, then the estimation error has the following upper bound as \(k \rightarrow \infty\): 
	\begin{gather}
		\|\tilde{\theta}_{k+1}\|^2 = O\left(\frac{\log \lambda_{\max}(k)}{\lambda_{\min}(k)}\right), \quad \text{a.s.}, 
	\end{gather}
\noindent where \(\tilde{\theta}_{k+1} = \theta - \theta_{k+1}\), \(\lambda_{\min}(k)\) and \(\lambda_{\max}(k)\) are the smallest and largest eigenvalues of the matrix \(P_0^{-1} + \sum\limits_{i=0}^{k} \phi_i \phi_i^{\mathrm{T}}\), respectively.
\end{theorem}

See Appendix \ref{Appendix proof} for the proof of Theorem \ref{mainThm}.

\begin{remark}
According to Theorem \ref{mainThm}, the estimates \(\{\theta_k, k \geq 0\}\) will converge to the true parameter \(\theta\) almost surely if
\begin{equation}\label{npe}
    \lim_{k \rightarrow \infty} \frac{\log \lambda_{\max}(k)}{\lambda_{\min}(k)} = 0, \quad \text{a.s.}
\end{equation} 
\noindent The condition \eqref{npe} is far weaker than the traditional PE condition, i.e., $\lambda_{\max}(k) = O\left(\lambda_{\min}(k)\right)$. \eqref{npe} aligns with the excitation condition derived under the \(\ell_1\)-norm optimization for saturated observation models (\cite{zheng2024l1}), and is known as the weakest excitation condition for the least squares estimates of stochastic linear regression models (\cite{lai1982least}). Moreover, the global convergence of the parameter estimates ensures that the projection operator will no longer be needed after some finite time, since the true parameter resides within the interior of the compact set \(\mathcal{K}\).
\end{remark}
Furthermore, to measure the deviation between the best predictor  \(\hat{y}_{k+1}^{*}\) defined in \eqref{predictc} and the adaptive predictor \(\hat{y}_{k+1}\) defined in \eqref{(8)} , we define the following regret:
\begin{equation}
	R_k = |\hat{y}_{k+1}^{*} - \hat{y}_{k+1}|.
\end{equation}
To formulate the separation condition between the true signal and the quantization thresholds, define
$
\zeta_i=\min_{1\le j\le m}\left|\phi_i^{\mathrm T}\theta-c_{ji}\right|.
$
The next theorem establishes the convergence of the average regret.

\begin{theorem} \label{regretThm} 
Suppose that Assumptions \ref{threshold}--\ref{symmetry} are satisfied. 
Assume further that the unquantized signal is asymptotically separated from the quantization thresholds in the sense that $
\lim_{\eta\downarrow 0}\limsup_{k\to\infty}
\frac{1}{k}\sum_{i=1}^k
\operatorname{I}_{\{\zeta_i\le \eta\}}
=0$ almost surely.
Then the average regret converges to zero, i.e.,
\begin{gather}\label{regrettheorem}
    \lim_{k \rightarrow \infty}\frac{1}{k}\sum_{i=1}^{k}R_i = 0, \quad \text{a.s.}
\end{gather}
\end{theorem}
The proof of Theorem \ref{regretThm} is given in Appendix \ref{Appendix proof}. 
\begin{remark}
     Theorem \ref{regretThm} demonstrates that the cumulative regret between the optimal predictor and the adaptive predictor increases at a rate of $o(k)$. Notably, The result \eqref{regrettheorem} does not need any excitation conditions, making it applicable to closed-loop control systems.
\end{remark}

\section{Application to Probation Decision Problems} \label{Experiment}

Developing technologies for data-driven sentencing assistance based on judicial data has attracted growing attention in China. Because sentencing is inherently tied to fairness and justice, the public is highly concerned about its normative aspects, which in turn imposes stringent requirements on the reliability of computational results (\cite{wang2022applications}). The adaptive identification algorithm developed in this paper is suitable for sentencing scenarios with non-i.i.d. data, as it imposes no such assumptions while offering theoretical guarantees of convergence. Since the probation decision (i.e., whether probation is granted) yields a binary outcome determined by multiple criminal circumstances specified in the sentencing guidelines, it provides a suitable setting for evaluating the performance of the ADA algorithm. Therefore, we collected 30{,}728 intentional injury cases from China Judgments Online, covering the years 2014 to 2024. Based on this dataset, we conducted an accuracy comparison between the proposed $\ell_{1}$-based algorithm and the $\ell_{2}$-based two-step quasi-Newton (TSQN) algorithm proposed in \cite{ZhangTAC2024}, which is also suitable for binary observations. The evaluation metric is defined as follows: 
\begin{gather}\label{sentencing_loss}
     \frac{1}{T}\sum_{k=1}^{T} \operatorname{I}_{\left[y_k = \hat{y}_k\right]},
\end{gather}
\noindent where \( T \) is the total number of sentencing cases.

The ADA algorithm adopts the following configuration: the noise sequence $\{\epsilon_k\}$ is independent and follows the distribution $\mathcal{N}(0,25)$. The convex compact set $\mathcal{K}$ is assumed to be the hypercube $[-1,1]^d$, i.e., each component lies within the interval $[-1,1]$. The threshold $c_{1k}=0.5$, $G_k = \sup\limits_{x \in \mathcal{K}} |\phi_k^{\mathrm{T}} x|$, $\bar{\mu}_k = \mu_k = 1$, and initialize the matrices as $\bar{P}_0 = P_0 = I$. The initial parameters $\bar{\theta}_0$ and $\theta_0$ are initialized at the zero vector. The hyperparameters of the TSQN algorithm are set the same for a fair comparison.

\begin{figure}[ht]
    \centering    \includegraphics[width=0.29\textwidth]{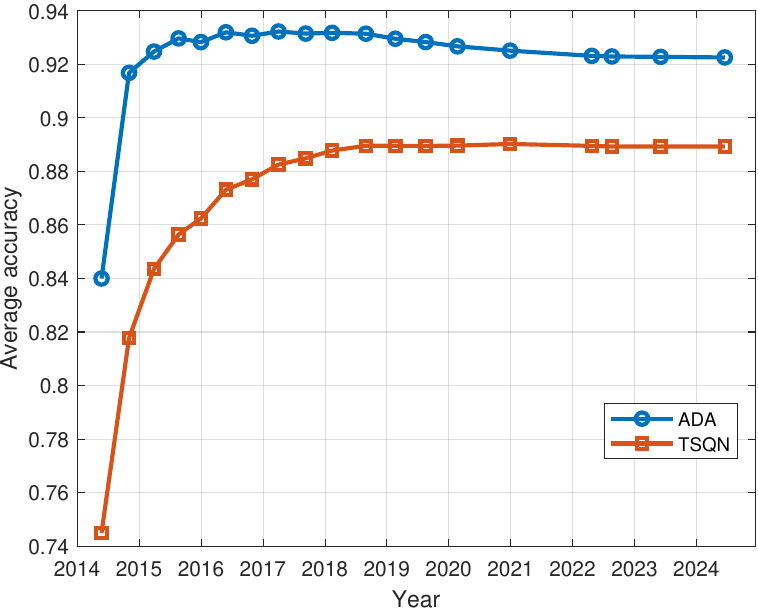}  
    \caption{Comparison of adaptive prediction accuracy.}
    \label{Accuracy}
\end{figure}

Fig.~\ref{Accuracy} shows that the ADA algorithm outperforms the TSQN algorithm in prediction accuracy as defined in \eqref{sentencing_loss}.

\section{Conclusion} \label{conclusion}

 In this paper, we have proposed a two-step adaptive algorithm based on $\ell_1$-norm loss function. Under non-PE data conditions, we have established both the global convergence of parameter estimator and the averaged regret of the adaptive algorithm. It is worth pointing out that our adaptive algorithm is motivated by real-world challenges including judicial sentencing problems, which in turn has been applied successfully to a basic classification problem in judicial sentencing  with real datasets. Future work may explore tailored adaptive algorithms for various model structures to better address diverse application scenarios.

\bibliography{ifacconf}             
                                                   







\appendix

 \section{Proof of The Theorems}\label{Appendix proof}
To complete the proof of Theorem \ref{mainThm}, we introduce the following notation:
\begin{gather}\label{barpsik}
\begin{aligned}
\bar{\psi}_k = & \mathbb{E}_k\left\{\operatorname{sgn}\left[y_{k+1}-S_k(\ptb)\right] \right\} \\
&- \left[1- F_{k+1}(c_{1k} - \ptb)\right]\operatorname{I}_{\left[\phi_k^{\mathrm{T}} \bar{\theta}_{k} \leq c_{1k}\right]} \\
& + F_{k+1}(c_{mk} - \ptb)\operatorname{I}_{\left[\phi_k^{\mathrm{T}}\bar{\theta}_{k} > c_{mk}\right]} \\
&+  \scalebox{0.675}{$  \sum\limits_{i=2}^{m} \left\{ F_{k+1}\left(c_{(i-1)k}-\phi_k^{\mathrm{T}}\bar{\theta}_{k}\right)   -\left[1-F_{k+1}\left(c_{ik}-\phi_k^{\mathrm{T}}\bar{\theta}_{k}\right)\right]\right\} \operatorname{I}_{ [c_{(i-1)k} < \phi_k^{\mathrm{T}}\bar{\theta}_{k} \leq c_{ik}]} $},
\end{aligned}
\end{gather}
\begin{gather}\label{psik}
\begin{aligned}
\psi_k =& \mathbb{E}_k\left\{\operatorname{sgn}\left[y_{k+1}-S_k(\pt)\right] \right\} \\
&- \left[1- F_{k+1}(c_{1k} - \pt)\right]\operatorname{I}_{\left[\phi_k^{\mathrm{T}} \theta_{k} \leq c_{1k}\right]} \\
& +F_{k+1}(c_{mk} - \pt)\operatorname{I}_{\left[\phi_k^{\mathrm{T}}\theta_{k} > c_{mk}\right]} \\
&+  \scalebox{0.675}{$ \sum\limits_{i=2}^{m} \left\{ F_{k+1}\left(c_{(i-1)k}-\phi_k^{\mathrm{T}}\theta_{k}\right)   -\left[1-F_{k+1}\left(c_{ik}-\phi_k^{\mathrm{T}}\theta_{k}\right)\right]\right\} \operatorname{I}_{ [c_{(i-1)k} < \phi_k^{\mathrm{T}}\theta_{k} \leq c_{ik}]} $},
\end{aligned}
\end{gather}
besides, 
\begin{gather}\label{barwk+1}
\scalebox{0.82}{$\begin{aligned}
   \bar{w}_{k+1} =& \operatorname{sgn} \left[y_{k+1} - S_k(\phi_k^{\mathrm{T}}\bar{\theta}_k)\right]
 - \mathbb{E}_k\left\{ \operatorname{sgn}\left[y_{k+1}  - S_k(\phi_k^{\mathrm{T}}\bar{\theta}_k)\right]\right\} ,
 \end{aligned}$}
\end{gather}
\begin{equation}\label{wk+1}
\scalebox{0.83}{$\begin{aligned}
w_{k+1} =  \operatorname{sgn}\left[y_{k+1} - S_k(\pt)\right] 
-\mathbb{E}_k\left\{\operatorname{sgn}\left[y_{k+1} - S_k(\pt)\right]\right\}.
\end{aligned}$}
\end{equation}

With Assumptions \ref{threshold}--\ref{symmetry}, one can deduce that  \(\{\bar{w}_{k}, \mathcal{F}_k, k \geq 0\}\) and \(\{w_{k}, \mathcal{F}_k, k \geq 0\}\) are  martingale difference sequences, i.e., \(\mathbb{E}_k\left[\bar{w}_{k+1}\right] =0\) and \(\mathbb{E}_k\left[w_{k+1}\right] =0\) for \(\forall k \geq 0\). Furthermore, one can see that \(\sup\limits_{k\geq 0}\mathbb{E}_k\left[|\bar{w}_{k+1}|^{2+\eta}\right] < \infty\) and \(\sup\limits_{k\geq 0}\mathbb{E}_k\left[|w_{k+1}|^{2+\eta}\right] < \infty\), where the constant \(\eta>0\).

In addition, we need the following lemmas to prove Theorem \ref{mainThm}.


\begin{lemma}\label{marting} (\cite{chen1991identification}). Let \(\left\{w_k, \mathcal{F}_k, k \geq 0 \right\}\) be a martingale difference sequence and \(\left\{f_k, \mathcal{F}_k, k \geq 0\right\}\) an adapted sequence. If \(\sup\limits_{k\geq 0} \mathbb{E}\left[\left|w_{k+1}\right|^\alpha \mid \mathcal{F}_k\right]<\infty, \text { a.s. }\), for some \(\alpha \in(0,2]\), then as \(k \rightarrow \infty\) :
\begin{gather}
\sum_{i=0}^k f_i w_{i+1}=O\left(s_k(\alpha) \log ^{\frac{1}{\alpha}+\eta}\left(s_k^\alpha(\alpha)+ \gamma\right)\right),~\text{a.s.}, ~\forall \eta>0,
\end{gather}
\noindent where \(s_k(\alpha)=\left(\sum\limits_{i=0}^k\left|f_i\right|^\alpha\right)^{\frac{1}{\alpha}}\) and the constant \( \gamma >0\) .
\end{lemma}

\begin{lemma}\label{lem1} 
(\cite{lai1982least}). Let \(X_1, X_2, \cdots\) be a sequence of vectors in \(\mathbb{R}^d~(d \geq 1)\) and let \(A_k=A_0+\sum\limits_{i=1}^k X_i X_i^\mathrm{T}\). Assume that \(A_0\) is nonsingular, then as \(k \rightarrow \infty\):
\begin{equation}
\sum_{k=1}^n \frac{X_k^T A_{k-1}^{-1} X_k}{1+X_k^T A_{k-1}^{-1} X_k} \leq \log \left(\left|A_n\right|\right)+\log \left(\left|A_0\right|\right) ,
\end{equation}
where \( |A_n| \) denotes the determinant of \( A_n \).
\end{lemma}

\begin{lemma}\label{lem2} (\cite{guo1995convergence}). Let \(X_1, X_2, \cdots\) be any bounded sequence of vectors in \(\mathbb{R}^d (d \geq 1)\). Denote \(A_k=A_0+ \sum\limits_{i=1}^k X_i X_i^T\) with the matrix \(A_0>0\), then we have
\begin{gather}	
\sum_{k=1}^{\infty}\left(X_k^T A_{k-1}^{-1} X_k\right)^2<\infty.
\end{gather}
\end{lemma}

\begin{lemma}
\label{l1}
Under Assumptions \ref{threshold}--\ref{symmetry}, \(|\bar{\psi}_k| \leq 2\) and \(|\psi_k| \leq 2\). 
\end{lemma}

\begin{pf}
We only present the proof of $|\bar{\psi}_k| \leq 2$, as the proof of $|\psi_k| \leq 2$ is completely analogous. In addition, \(\mathbb{P}_{k+1}(\cdot)\) denotes the conditional probability function of \(y_{k+1}\) given \(\mathcal{F}_k\) in this proof. 

The proof of \(|\bar{\psi}_k| \leq 2\) is given in different cases below.

If \(\ptb \leq c_{1k}\), then \(S_k(\ptb) = 1\). Noticing that \(\mathbb{P}_{k+1}(y_{k+1} < 1)=0\), one can deduce that
\begin{equation}
    \begin{aligned}
	\bar{\psi}_{k} =& \mathbb{E}_k\left\{\operatorname{sgn}\left[y_{k+1}^c - 1\right]\right\}-\left[1-F_{k+1}(c_{1k}-\phi_k^{\mathrm{T}}\bar{\theta}_{k})\right]\\
	=&\mathbb{P}_{k+1}(y_{k+1}> 1)-\left[1-F_{k+1}(c_{1k}-\phi_k^{\mathrm{T}}\bar{\theta}_{k})\right]	\\
	 =& F_{k+1}(c_{1k}- \phi_k^{\mathrm{T}}\bar{\theta}_{k})-F_{k+1}(c_{1k}-\phi_k^{\mathrm{T}}\theta),		
    \end{aligned}
\end{equation}
\noindent where the last equality holds since the density function \(f_{k+1}(\cdot)\) is continuous. Then we can see that \(|\bar{\psi}_k| \leq 1\) in this case.
 
As for \(\ptb > c_{mk}\), following a similar analysis to the case of \(\ptb \leq c_{1k}\),  we have \(|\bar{\psi}_k| \leq 1\) in this case.

If \( c_{(i-1)k}<\ptb \leq c_{ik} \) (\(i =2, \cdots, m\)),  one can deduce that
\begin{equation}
    \begin{aligned}
	\bar{\psi}_k =& \mathbb{E}_k\left\{\operatorname{sgn}\left[y_{k+1}-S_k(\ptb)\right]\right\}  + F_{k+1}\left(c_{(i-1)k} -\phi_k^{\mathrm{T}}\bar{\theta}_{k}\right) \\
    & -\left[1-F_{k+1}\left(c_{ik}-\phi_k^{\mathrm{T}}\bar{\theta}_{k}\right)\right]  \\
	  =& \left[F_{k+1}\left(c_{(i-1)k}-\phi_k^{\mathrm{T}}\bar{\theta}_{k}\right)-F_{k+1}\left(c_{(i-1)k}-\phi_k^{\mathrm{T}}\theta\right) \right] \\
      &+\left[F_{k+1}\left(c_{ik}-\phi_k^{\mathrm{T}}\bar{\theta}_{k}\right)-F_{k+1}\left(c_{ik}-\phi_k^{\mathrm{T}}\theta\right) \right],
    \end{aligned}
\end{equation}
\noindent where the last equality holds due to the continuity of the density function \(f_{k+1}(\cdot)\). Therefore, we have \(|\bar{\psi}_k| \leq 2\) in this case. \hfill$\square$
\end{pf}

\begin{lemma} \label{critical}
Under Assumptions \ref{threshold}--\ref{symmetry},	\(|\psi_k - \beta_k \phi_k^{\mathrm{T}}\tilde{\theta}_k| =  O\left(|\phi_k^{\mathrm{T}}\tilde{\bar{\theta}}_k|\right),\) \text{a.s.}, where \(\tilde{\bar{\theta}}_k =  \theta - \bar{\theta}_k\).
\end{lemma}
\begin{pf}
The proof is completed under different cases below.

If \(\phi_k^{\mathrm{T}}\theta_{k} \leq c_{1k}\), we have 
\begin{equation}
    \begin{aligned}
	&\psi_k - \left[F_{k+1}(c_{1k} - \pt)-F_{k+1}(c_{1k} - \ptb)\right]\\ 
	=&\psi_k - \beta_k \phi_k^{\mathrm{T}}(\bar{\theta}_k - \theta_k) \\
    \end{aligned}
\end{equation}
\noindent noticing that \(\psi_k = F_{k+1}(c_{1k} - \pt) - F_{k+1}(c_{1k} - \phi_k^{\mathrm{T}}\theta) \), then \(\psi_k - \beta_k \phi_k^{\mathrm{T}}\tilde{\theta}_k = \left(f_{k+1}(z_1)- \beta_k\right)\phi_k^{\mathrm{T}}\tilde{\bar{\theta}}_k\), where the value \(z_1\) is between \( c_{1k} - \ptb\) and \(c_{1k} - \phi_k^{\mathrm{T}}\theta\) according to the Mean Value Theorem.	

If \(\phi_k^{\mathrm{T}}\theta_{k}>c_{mk}\), following the analogous analysis to the case of \(\phi_k^{\mathrm{T}}\theta_{k} < c_{1k}\), we also have \(\psi_k - \beta_k \phi_k^{\mathrm{T}}\tilde{\theta}_k = \left(f_{k+1}(z_2) - \beta_k\right)\phi_k^{\mathrm{T}}\tilde{\bar{\theta}}_k\), where the \(z_2\) takes values between \( c_{mk} - \ptb\) and \(c_{mk} - \phi_k^{\mathrm{T}}\theta\) according to the Mean Value Theorem.

If \( c_{(i-1)k} < \phi_k^{\mathrm{T}}\theta_{k}< c_{ik}\) \((i=2, \cdots, m)\), the following fact holds:	
\begin{equation}
    \begin{aligned}
        &\psi_k - \left[F_{k+1}\left(c_{(i-1)k}-\phi_k^{\mathrm{T}}\theta_{k}\right)  - F_{k+1}\left(c_{(i-1)k}-\phi_k^{\mathrm{T}}\bar{\theta}_k\right) \right.\\  
         & \left.+ F_{k+1}\left(c_{ik}-\phi_k^{\mathrm{T}}\theta_{k}\right)-F_{k+1}\left(c_{ik}-\phi_k^{\mathrm{T}}\bar{\theta}_k\right) \right]\\
     =& \psi_k - \beta_k \phi_k^{\mathrm{T}}\tilde{\theta}_k + \beta_k \phi_k^{\mathrm{T}}\tilde{\bar{\theta}}_k,
    \end{aligned}
\end{equation}
\noindent since $\psi_k = \left[F_{k+1}\left(c_{(i-1)k}-\phi_k^{\mathrm{T}}\theta_{k}\right)-F_{k+1}\left(c_{(i-1)k}-\phi_k^{\mathrm{T}}\theta\right) \right]$ $ + \left[F_{k+1}\left(c_{ik}-\phi_k^{\mathrm{T}}\theta_{k}\right)-F_{k+1}\left(c_{ik}-\phi_k^{\mathrm{T}}\theta\right) \right]$, so \(\psi_k -\beta_k\phi_k^{\mathrm{T}}\tilde{\theta}_k \) \( = \left[f_{k+1}(z_3) + f_{k+1}(z_4)- \beta_k\right]\phi_k^{\mathrm{T}}\tilde{\bar{\theta}}_k\), where \(z_3\) takes values between \(c_{(i-1)k} - \ptb \) and \(c_{(i-1)k} - \phi_k^{\mathrm{T}}\theta\) and \(z_4\) takes values between \(c_{ik} - \ptb \) and \(c_{ik} - \phi_k^{\mathrm{T}}\theta\) according to the Mean Value Theorem.
	
Under Assumption \ref{symmetry}, \(f_{k+1}(z_i) (i=1, 2, 3, 4)\) and \(\beta_k\) are bounded almost surely. Hence Lemma \ref{critical} holds.  \hfill$\square$
\end{pf}

\begin{lemma}\label{lemma3}
Under Assumptions \ref{threshold}--\ref{symmetry}, \(|\bar{\psi}_k| \geq \bar{\beta}_k |\phi_k^{\mathrm{T}}\tilde{\bar{\theta}}_k|\) and \(\phi_k^{\mathrm{T}}\tilde{\bar{\theta}}_k\bar{\psi}_k\geq 0\).
\end{lemma}

\begin{pf}
This proof is completed under different cases below.

If \(\ptb \leq c_{1k}\), according to the proof of Lemma \ref{l1},
\begin{equation}
    \begin{aligned}
	\bar{\psi}_{k} &= F_{k+1}(c_{1k} - \phi_k^{\mathrm{T}}\bar{\theta}_{k})-F_{k+1}(c_{1k} -\phi_k^{\mathrm{T}}\theta)	\\	
      &= f_{k+1}(e_1)\phi_k^{\mathrm{T}}\tilde{\bar{\theta}}_k,
    \end{aligned}
\end{equation}
\noindent where the \(e_1\) is between \( c_{1k} - \ptb\) and \(c_{1k} - \phi_k^{\mathrm{T}}\theta\) according to the Mean Value Theorem. Under Assumption \ref{symmetry}, one can deduce that \(f_{k+1}(e_1)\geq \bar{\beta}_k\), hence \(|\bar{\psi}_k| \geq \bar{\beta}_k |\phi_k^{\mathrm{T}}\tilde{\bar{\theta}}_k|\). Besides, \(\phi_k^{\mathrm{T}}\tilde{\bar{\theta}}_k\bar{\psi}_k = f_{k+1}(e_1) (\phi_k^{\mathrm{T}}\tilde{\bar{\theta}}_k)^2\geq 0\).

As for the case where \(\phi_k^{\mathrm{T}}\bar{\theta}_{k} > c_{mk}\), following the same analysis above, we have similar results.

If \( c_{(i-1)k} < \phi_k^{\mathrm{T}}\theta_{k}< c_{ik}\) \((i=2, \cdots, m)\), combing Assumption \ref{symmetry}, we know that
\begin{equation}
    \begin{aligned}
	\bar{\psi}_{k} &= \left[F_{k+1}\left(c_{(i-1)k}-\phi_k^{\mathrm{T}}\bar{\theta}_{k}\right)-F_{k+1}\left(c_{(i-1)k}-\phi_k^{\mathrm{T}}\theta\right) \right] \\
    &+\left[F_{k+1}\left(c_{ik}-\phi_k^{\mathrm{T}}\bar{\theta}_{k}\right)-F_{k+1}\left(c_{ik}-\phi_k^{\mathrm{T}}\theta\right) \right]	\\	
    &= \left[f_{k+1}(e_3) + f_{k+1}(e_4) \right]\phi_k^{\mathrm{T}}\tilde{\bar{\theta}}_{k},
    \end{aligned}
\end{equation}
\noindent where the \(e_3\) is between \(c_{(i-1)k}-\phi_k^{\mathrm{T}}\bar{\theta}_{k}\) and \(c_{(i-1)k}-\phi_k^{\mathrm{T}}\theta\) and \(e_4\) is between \(c_{ik}-\phi_k^{\mathrm{T}}\bar{\theta}_{k}\) and \(c_{ik}-\phi_k^{\mathrm{T}}\theta\) according to the Mean Value Theorem. Under Assumption \ref{symmetry}, we know that \(f_{k+1}(e_3)\geq \bar{\beta}_k \) and \(f_{k+1}(e_4)\geq \bar{\beta}_k \), hence \(|\bar{\psi}_k| \geq 2\bar{\beta}_k |\phi_k^{\mathrm{T}}\tilde{\bar{\theta}}_k|\). Besides, \(\phi_k^{\mathrm{T}}\tilde{\bar{\theta}}_k\bar{\psi}_k = \left[f_{k+1}(e_3)+f_{k+1}(e_4)\right] (\phi_k^{\mathrm{T}}\tilde{\bar{\theta}}_k)^2\geq 0\).
\end{pf}

\begin{lemma}\label{lemma1}
Let Assumptions \ref{threshold}--\ref{symmetry} be satisfied, then the parameter estimate \(\bar{\theta}_{k+1}\) given by Algorithm \ref{algorithm1} has the following property almost surely as \(k \rightarrow \infty\):
	\begin{gather} \label{desired}
        \begin{aligned}
       \tilde{\bar{\theta}}_{k+1}^{\mathrm{T}} \bar{P}_{k+1}^{-1} \tilde{\bar{\theta}}_{k+1}+ \sum_{i=0}^k \bar{\mu}_i^{-1}\bar{\beta}_i (\phi_i^{\mathrm{T}}\tilde{\bar{\theta}}_i)^2 =O\left(\log \lambda_{\max }\left(k\right)\right),
       \end{aligned}
    \end{gather}
    where \(\lambda_{\max}(k)\) is the largest eigenvalue of the matrix \(\bar{P_0}^{-1} + \sum\limits_{i=0}^{k} \phi_i \phi_i^{\mathrm{T}}\).
\end{lemma}

\begin{pf}
Inspired by the classical recursive least square algorithm for linear stochastic models in \cite{guo1995convergence}, the \(\ell_2\)-norm based methods in \cite{ZhangTAC2024} and the \(\ell_1\)-norm based methods in \cite{zheng2024l1}, we choose the following stochastic Lyapunov function:
\begin{gather}\label{Vk}
    \bar{V}_{k+1} = \tilde{\bar{\theta}}_{k+1}^{\mathrm{T}}\bar{P}_{k+1}^{-1}\tilde{\bar{\theta}}_{k+1}.
\end{gather}
  
According to the definition of \(\bar{P}_{k+1}\) of Algorithm \ref{algorithm1} and the well-known matrix inversion formula (see e.g., \cite{guo2020tima}), we have
\begin{gather} \label{Pni}
    \bar{P}_{k+1}^{-1} = \bar{P}_{k}^{-1} + \bar{\mu}_k^{-1}\bar{\beta}_k  \phi_k \phi_k^{\mathrm{T}}.
\end{gather}
Moreover, multiplying the left side of \eqref{Pni} by \(\bar{a}_k\phi_k^{\mathrm{T}}\bar{P}_k\), we have
\begin{equation} \label{leftP}
    \begin{aligned}
        \bar{a}_k\phi_k^{\mathrm{T}}\bar{P}_k \bar{P}_{k+1}^{-1} 
        =\bar{a}_k\phi_k^{\mathrm{T}} (I + \bar{\mu}_k^{-1}\bar{\beta}_k \bar{P}_k\phi_k \phi_k^{\mathrm{T}}) = \bar{\mu}_k^{-1} \phi_k^{\mathrm{T}}.
    \end{aligned}
\end{equation}

Therefore, by Lemma 2 in \cite{zheng2024l1}, \eqref{Vk}, \eqref{Pni} and \eqref{leftP} , one can deduce that
\begin{gather}\label{barVk+1}
    \begin{aligned}
	\bar{V}_{k+1} \leq& \left[\tilde{\bar{\theta}}_k - \bar{a}_k\bar{P}_k\phi_k\left(\bar{\psi}_k + \bar{w}_{k+1}\right)\right]^{\mathrm{T}}\bar{P}_{k+1}^{-1}\\
     & \left[\tilde{\bar{\theta}}_k - \bar{a}_k\bar{P}_k\phi_k\left(\bar{\psi}_k + \bar{w}_{k+1}\right)\right]\\
        \leq& \scalebox{0.93}{$ \tilde{\bar{\theta}}_{k}^{\mathrm{T}}\bar{P}_{k+1}^{-1}\tilde{\bar{\theta}}_{k} - 2\bar{a}_k\phi_k^{\mathrm{T}}\bar{P}_k\bar{P}_{k+1}^{-1}\tilde{\bar{\theta}}_{k}\bar{\psi}_k 
	   +\bar{a}_k^2  \phi_k^{\mathrm{T}} \bar{P}_k \bar{P}_{k+1}^{-1}\bar{P}_k\phi_k \bar{\psi}_k^2 $}\\
	   & + \scalebox{0.95}{$ 2\bar{a}_k^2 \phi_k^{\mathrm{T}} \bar{P}_k \bar{P}_{k+1}^{-1} \bar{P}_k\phi_k \bar{\psi}_k \bar{w}_{k+1}
	   -2\bar{a}_k\phi_k^{\mathrm{T}}\bar{P}_k \bar{P}_{k+1}^{-1}\tilde{\bar{\theta}}_k \bar{w}_{k+1} $}\\
	   & +\bar{a}_k^2 \phi_k^{\mathrm{T}}\bar{P}_k \bar{P}_{k+1}^{-1}\bar{P}_k\phi_k \bar{w}_{k+1}^2 \\
	   = & \scalebox{0.93}{$ \bar{V}_k + \bar{\mu}_k^{-1} \bar{\beta}_k (\phi_k^{\mathrm{T}}\tilde{\bar{\theta}}_k)^2 - 2\bar{\mu}_k^{-1}\phi_k^{\mathrm{T}}\tilde{\bar{\theta}}_k \bar{\psi}_k 
	   +\bar{\mu}_k^{-1} \bar{a}_k \phi_k^{\mathrm{T}}\bar{P}_k\phi_k \bar{\psi}_k^2 $}\\
          & + 2\bar{\mu}_k^{-1}\bar{a}_k  \phi_k^{\mathrm{T}}\bar{P}_k\phi_k \bar{\psi}_k \bar{w}_{k+1} 
	   -2 \bar{\mu}_k^{-1}\phi_k^{\mathrm{T}}\tilde{\theta}_k \bar{w}_{k+1} \\
       & + \bar{\mu}_k^{-1}\bar{a}_k\phi_k^{\mathrm{T}}\bar{P}_k\phi_k \bar{w}_{k+1}^2, \\
    \end{aligned}
\end{gather}
where \(\bar{\psi}_k\) and \(\bar{w}_{k+1}\) are defined in \eqref{barpsik} and \eqref{barwk+1}, respectively. Moreover, according to Lemma \ref{lemma3} and \(\bar{\mu}_k > 0\), we have
\begin{equation}\label{decreasekey}
\begin{aligned}
    &\bar{\mu}_k^{-1} \bar{\beta}_k (\phi_k^{\mathrm{T}}\tilde{\bar{\theta}}_k)^2 - 2\bar{\mu}_k^{-1}\phi_k^{\mathrm{T}}\tilde{\bar{\theta}}_k \bar{\psi}_k 
    \leq - \bar{\mu}_k^{-1} \bar{\beta}_k (\phi_k^{\mathrm{T}}\tilde{\bar{\theta}}_k)^2.
\end{aligned}
\end{equation}

Summing up both sides of \eqref{barVk+1} from 0 to \(n\), and noticing Lemma \ref{l1}, \eqref{decreasekey},  we have
\begin{gather} \label{barVc}
\begin{aligned}
    \bar{V}_{n+1} \leq& \bar{V}_0 - \sum_{k=0}^{n}\bar{\mu}_k^{-1} \bar{\beta}_k (\phi_k^{\mathrm{T}}\tilde{\bar{\theta}}_k)^2   
     + \frac{4}{h}\sum_{k=0}^{n}\bar{\mu}_k^{-1} \bar{a}_k \bar{\beta}_k\phi_k^{\mathrm{T}}\bar{P}_k\phi_k  \\
    &\scalebox{0.95}{$ + 2\sum\limits_{k=0}^{n}\bar{\mu}_k^{-1} \bar{a}_k  \phi_k^{\mathrm{T}}\bar{P}_k\phi_k \bar{\psi}_k \bar{w}_{k+1}  
    -2\sum\limits_{k=0}^{n}\bar{\mu}_k^{-1}\phi_k^{\mathrm{T}}\tilde{\theta}_k \bar{w}_{k+1} $}\\
    &+ \frac{1}{h}\sum_{k=0}^{n}\bar{\mu}_k^{-1}\bar{a}_k \bar{\beta}_k\phi_k^{\mathrm{T}}\bar{P}_k\phi_k \bar{w}_{k+1}^2,~ \text{a.s.},
\end{aligned}
\end{gather}
\noindent where \(h= \inf\limits_{|x|\leq G+ \Lambda_c,\ k\geq 0} f_{k+1}(x)>0\).

Now let us analyze the RHS of \eqref{barVc} term by term.

First of all, for the third term on the RHS of \eqref{barVc}, we let \(X_{k}= \sqrt{\bar{\mu}_{k}^{-1}\bar{\beta}_k}\phi_k\) in Lemma \ref{lem1}, yielding
\begin{gather}\label{step1begin}
\begin{aligned}
    &\sum_{k=0}^{n}\bar{\mu}_k^{-1}\bar{a}_k \bar{\beta}_k\phi_k^{\mathrm{T}}\bar{P}_k\phi_k	
    = O\left(\log\lambda_{\max}(n) \right),
\end{aligned}
\end{gather}
\noindent where \(\bar{\mu}_{\inf} = \inf\limits_{k\geq0} \bar{\mu}_k > 0\), and the last equality holds due to the boundedness of \(\{\bar{\mu}_k, k \geq 0\}\) and \(\{\bar{\beta}_k, k \geq 0\}\).

As for the fourth term on the RHS of \eqref{barVc}, applying Lemma \ref{marting} and the condition \(\sup\limits_{k\geq 0}\mathbb{E}_k\left[|\bar{w}_{k+1}|^{2}\right] < \infty\), we have
\begin{equation}\label{reffirst}
    \begin{aligned}
        &\sum_{k=0}^{n}\bar{\mu}_k^{-1} \bar{a}_k  \phi_k^{\mathrm{T}}\bar{P}_k\phi_k \bar{\psi}_k \bar{w}_{k+1} \\
        = &o(\sum_{k=0}^{n}(\bar{\mu}_k^{-1} \bar{a}_k \bar{\beta}_k \phi_k^{\mathrm{T}}\bar{P}_k\phi_k )^2) + O(1),~\text{a.s.},
    \end{aligned}
\end{equation}
where  the  equality holds due to Lemma \ref{l1} and the fact that \(\{\bar{\mu}_k, k \geq 0\}\) and \(\{\bar{\beta}_k, k \geq 0\}\) are uniformly bounded for \(k \geq 0\). Let \(X_{k}= \sqrt{\bar{\mu}_{k}^{-1}\bar{\beta}_k}\phi_k\) in Lemma \ref{lem2}, one can deduce that
\begin{equation} \label{bound39}
    \sum_{k=0}^{n}(\bar{\mu}_k^{-1} \bar{a}_k \bar{\beta}_k \phi_k^{\mathrm{T}}\bar{P}_k\phi_k )^2 = O(1),
\end{equation}
since \(\{\bar{a}_k, k \geq 0\}\), \(\{\bar{\mu}_k, k \geq 0\}\) and \(\{\bar{\beta}_k, k \geq 0\}\) are uniformly bounded for \(k \geq 0\). Therefore,
\begin{equation} \label{boundwpsi}
    \sum_{k=0}^{n}\bar{\mu}_k^{-1} \bar{a}_k  \phi_k^{\mathrm{T}}\bar{P}_k\phi_k \bar{\psi}_k \bar{w}_{k+1} = O(1),~\text{a.s.}
\end{equation}
Similarly, by Lemma \ref{marting}, we have
\begin{equation} \label{jiaocha}
    \sum_{k=0}^{n}\bar{\mu}_k^{-1}\phi_k^{\mathrm{T}}\tilde{\theta}_k \bar{w}_{k+1} = o(\sum_{k=0}^{n}\bar{\mu}_k^{-1} \bar{\beta}_k (\phi_k^{\mathrm{T}}\tilde{\bar{\theta}}_k)^2) + O(1), ~\text{a.s.},
\end{equation}
where we use the fact that \(\{\bar{\mu}_k, k \geq 0\}\) and \(\{\bar{\beta}_k, k \geq 0\}\) are uniformly bounded for \(k \geq 0\).

From the definition of \(\bar{w}_{k+1}\) in \eqref{barwk+1}, the following fact holds for any \(\tau \in (0,2]\),
\begin{gather}\label{bounduse}
	\sup_{k\geq 0} \mathbb{E}_k \left[ |\bar{w}_{k+1}^2 - \mathbb{E}_k[\bar{w}_{k+1}^2]|^{\tau}\right] < \infty,\quad  \text{a.s.}
\end{gather}
So for the last term on the RHS of \eqref{barVc}, using \eqref{bound39}, \eqref{bounduse} and Lemma \ref{marting}, we have
\begin{equation} \label{finalterm}
    \begin{aligned}
        &\sum_{k=0}^{n}\bar{\mu}_k^{-1}\bar{a}_k \bar{\beta}_k\phi_k^{\mathrm{T}}\bar{P}_k\phi_k \bar{w}_{k+1}^2 \\
        &= \sum_{k=0}^{n}\bar{\mu}_k^{-1}\bar{a}_k \bar{\beta}_k\phi_k^{\mathrm{T}}\bar{P}_k\phi_k (\bar{w}_{k+1}^2 - \mathbb{E}_k[\bar{w}_{k+1}^2]) \\
        &+ \sum_{k=0}^{n}\bar{\mu}_k^{-1}\bar{a}_k \bar{\beta}_k\phi_k^{\mathrm{T}}\bar{P}_k\phi_k \mathbb{E}_k[\bar{w}_{k+1}^2] \\
        & = O\left(\log\lambda_{\max}(n)\right), ~\text{a.s.},
    \end{aligned}
\end{equation}
where we use fact that \(\{\bar{w}_{k+1}^2 - \mathbb{E}_k[\bar{w}_{k+1}^2], k\geq 0\}\) is a martingale difference sequence and \(\sum\limits_{k=0}^{n}(\bar{\mu}_k^{-1}\bar{a}_k \bar{\beta}_k\phi_k^{\mathrm{T}}\bar{P}_k\phi_k)^2= O(1)\) almost surely by Lemma \ref{lem2}. Moreover, the last equality holds because of \eqref{step1begin} and the boundedness of \(\mathbb{E}_k\left[\bar{w}_{k+1}^2\right] \).

Now, by combing \eqref{barVc}, \eqref{step1begin}, \eqref{boundwpsi}, \eqref{jiaocha} and \eqref{finalterm}, we obtain the result \eqref{desired}.   \hfill$\square$
\end{pf}

\begin{lemma}\label{keylem}
Under Assumptions \ref{threshold}--\ref{symmetry}, the parameter estimate \(\theta_{k+1}\) given by Algorithm \ref{algorithm1} has the following property almost surely as \(k \rightarrow \infty\):
	\begin{gather}
        \begin{aligned}
       \tilde{\theta}_{k+1}^T P_{k+1}^{-1} \tilde{\theta}_{k+1}+ \sum_{i=0}^k \mu_i^{-1}\beta^{-1}_i \psi_i^2 =O\left(\log \lambda_{\max }\left(k\right)\right),
       \end{aligned}
    \end{gather}
    where \(\lambda_{\max}(k)\) is the largest eigenvalue of the matrix \(P_0^{-1} + \sum\limits_{i=0}^{k} \phi_i \phi_i^{\mathrm{T}}\).
\end{lemma}

\begin{pf}
We consider the following similar Lyapunov function:   
\begin{equation} \label{useful1}
    V_{k+1} = \tilde{\theta}_{k+1}^{\mathrm{T}}P_{k+1}^{-1}\tilde{\theta}_{k+1}.
\end{equation}


Moreover, note that 
\begin{equation} \label{square1}
\begin{aligned}
 &\mu_k^{-1} \beta_k (\phi_k^{\mathrm{T}}\tilde{\theta}_k)^2 - 2\mu_k^{-1}\phi_k^{\mathrm{T}}\tilde{\theta}_k \psi_k \\
=&  \mu_k^{-1}\beta_k^{-1}( \psi_k - \beta_k\phi_k^{\mathrm{T}}\tilde{\theta}_k)^2 - \mu_k^{-1} \beta_k^{-1} \psi_k^2, 
\end{aligned}
\end{equation}
besides, noticing the definition of \(a_k\) in Algorithm \ref{algorithm1},
\begin{gather}\label{square2}
    \begin{aligned}
	&\mu_k^{-1}a_k \phi_k^{\mathrm{T}}P_k\phi_k \psi_k w_{k+1} - \mu_k^{-1}\phi_k^{\mathrm{T}}\tilde{\theta}_kw_{k+1} \\
	=& \mu_k^{-1} \beta_k^{-1}(\psi_k - \beta_k\phi_k^{\mathrm{T}}\tilde{\theta}_k)w_{k+1}
         -  \beta_k^{-1}a_k\psi_k w_{k+1}.
    \end{aligned}
\end{gather}

Therefore, following similar analysis of \eqref{barVc}, by \eqref{square1} and \eqref{square2}, we have
\begin{gather} \label{Vc}
\begin{aligned}
    V_{n+1} \leq& V_0 - \sum_{k=0}^{n}\mu_k^{-1} \beta_k^{-1} \psi_k^2   
    + \sum_{k=0}^{n}\mu_k^{-1}\beta_k^{-1}( \psi_k - \beta_k\phi_k^{\mathrm{T}}\tilde{\theta}_k)^2  \\
     & \scalebox{0.85}{$ + \frac{4}{h}\sum\limits_{k=0}^{n}\mu_k^{-1}a_k\beta_k\phi_k^{\mathrm{T}}P_k\phi_k  
    + 2\sum\limits_{k=0}^{n}\mu_k^{-1}\beta_k^{-1}(\psi_k - \beta_k\phi_k^{\mathrm{T}}\tilde{\theta}_k)w_{k+1} $} \\ 
    & \scalebox{0.89}{$  -2\sum\limits_{k=0}^{n}\beta_k^{-1}a_k\psi_kw_{k+1} 
    + \frac{1}{h}\sum\limits_{k=0}^{n}\mu_k^{-1}a_k\beta_k \phi_k^{\mathrm{T}}P_k\phi_kw_{k+1}^2 $},~ \text{a.s.} 
\end{aligned}
\end{gather}

According to Lemmas \ref{critical} and \ref{lemma1}, one can deduce that \(\sum\limits_{k=0}^{n}\mu_k^{-1}\beta_k^{-1}( \psi_k - \beta_k\phi_k^{\mathrm{T}}\tilde{\theta}_k)^2 = O\left(\log \lambda_{\max }\left(n\right)\right)\).
The analysis of the other terms on the RHS of \eqref{Vc} follows the completely analogous way in the proof of Lemma \ref{lemma1}. Then one can see that Lemma \ref{keylem} holds. \hfill$\square$
\end{pf}

We are now in a position to prove Theorem \ref{mainThm}.

\textbf{Proof of Theorem \ref{mainThm}.} 
 With  Lemma \ref{keylem} and \eqref{useful1}, one can know that
\begin{gather} \label{proofofTHM}
V_{n+1} \geq c_0\lambda_{\min}(k)\| \tilde{\theta}_{n+1}\|^2, ~\text{a.s.},
\end{gather}
\noindent where \(c_0=\min\{1, \inf\limits_{k\geq 0}(\mu_k^{-1}\beta_k)\}\) is positive since the infimum of \(\{\mu_k, k \geq 0\}\)  and \(\{\beta_k, k \geq 0\}\) are greater than 0. Therefore, Theorem \ref{mainThm} follows immediately from Lemma \ref{keylem} and \eqref{proofofTHM}. \hfill$\square$

Now we are in a position to prove Theorem \ref{regretThm}. Before that, we need to introduce an important lemma regarding the convergence of series, i.e., the Koopman-von Neumann Theorem.

\begin{lemma} (\cite{guo2020tima}). \label{von neumann}
    Let \(\{a_k, k \geq 0\}\) be a bounded non-negative sequence. A necessary and sufficient condition for the limit \(\lim\limits_{n \to \infty} \frac{1}{n} \sum\limits_{k=0}^{n-1} a_k = 0\) to hold is that there exists a set \(E \subset \mathbb{Z}_{+}\) with density zero (i.e., \(\lim\limits_{n \to \infty} \frac{1}{n} \sum\limits_{k=0}^{n-1} I_{[k \in E]} = 0\)), such that \(\lim\limits_{\substack{n \to \infty \\ \ n \notin E}} a_n = 0\).
\end{lemma}

Now we are ready to prove Theorem \ref{regretThm}.

\noindent\textbf{Proof of Theorem \ref{regretThm}.}
By Lemma \ref{keylem}, and noting that $\|\phi_k\|$ is uniformly bounded, we have
\[
\frac{1}{k}\sum_{i=1}^k \mu_i^{-1}\beta_i^{-1}\psi_i^2
=
O\!\left(\frac{\log\lambda_{\max}(k)}{k}\right),\quad \text{a.s.}
\]
Moreover, $
\lim_{k\to\infty}\frac{\log\lambda_{\max}(k)}{k}=0$ almost surely. Hence,
\begin{equation}\label{regret_pf_1}
\lim_{k\to\infty}
\frac{1}{k}\sum_{i=1}^k \mu_i^{-1}\beta_i^{-1}\psi_i^2
=0,\quad \text{a.s.}
\end{equation}
By Lemma \ref{critical}, and noting that there exists a constant
$\underline{\beta}>0$ such that $\beta_i\geq \underline{\beta}$, we have
\[
(\phi_i^{\mathrm T}\tilde{\theta}_i)^2
\leq
M_\psi \psi_i^2
+
M_\psi(\phi_i^{\mathrm T}\tilde{\bar{\theta}}_i)^2,
\quad \text{a.s.}
\]
where $M_\psi>0$ is a constant independent of $i$. Together with Lemmas
\ref{lemma1} and \ref{keylem}, and the boundedness assumptions on
$\{\mu_i\}$, $\{\beta_i^{-1}\}$, $\{\bar{\mu}_i\}$, and
$\{\bar{\beta}_i^{-1}\}$, we further obtain
\begin{equation}\label{avgphierr}
\lim_{k\to\infty}
\frac{1}{k}\sum_{i=1}^k
(\phi_i^{\mathrm T}\tilde{\theta}_i)^2
=0,\quad \text{a.s.}
\end{equation}
Therefore, by the Koopman-von Neumann theorem (see, e.g., \cite{guo2020tima}), and by the uniform boundedness of $\phi_i$ and $\tilde{\theta}_i$, there exists a set
$E\subset\mathbb Z_+$ with asymptotic density zero such that
\begin{equation}\label{subseqphierr}
\lim_{\substack{k\to\infty\\ k\notin E}}
|\phi_k^{\mathrm T}\tilde{\theta}_k|
=0,\quad \text{a.s.}
\end{equation}

Since $
\lim_{k\to\infty}
\frac{1}{k}\sum_{i=1}^{k}\operatorname{I}_{\{i\in E\}}
=0 $
and $|R_i|\leq m<\infty$, we have
\begin{equation}\label{first0}
\lim_{k\to\infty}
\frac{1}{k}
\sum_{\substack{1\leq i\leq k\\ i\in E}}
R_i
=0,\quad \text{a.s.}
\end{equation}

For the term over $i\notin E$, introduce the event
\begin{equation}
\mathbb W_k:
\left[
\phi_k^{\mathrm T}\theta-|\phi_k^{\mathrm T}\tilde{\theta}_k|,
\,
\phi_k^{\mathrm T}\theta+|\phi_k^{\mathrm T}\tilde{\theta}_k|
\right]
\cap\{c_{1k},\dots,c_{mk}\}
=\emptyset .
\end{equation}
On $\mathbb W_k$, the true and estimated unquantized signals lie in the same
quantization interval. Hence,
\begin{equation}\label{Rk0}
R_k=|\hat y_{k+1}^{*}-\hat y_{k+1}|=0.
\end{equation}
Moreover, $\mathbb W_k$ can be equivalently written as
\begin{equation}
\mathbb W_k
=
\bigcap_{j=1}^{m}
\left\{
\left|\phi_k^{\mathrm T}\theta-c_{jk}\right|
>
|\phi_k^{\mathrm T}\tilde{\theta}_k|
\right\}.
\end{equation}
Define $
E_{\mathbb W}:=\{k:\mathbb W_k\text{ occurs}\}. $
Then
\begin{equation}\label{criticalregret}
\frac{1}{k}
\sum_{\substack{1\leq i\leq k\\ i\notin E}}R_i
=
\frac{1}{k}
\sum_{\substack{1\leq i\leq k\\ i\notin E\\ i\in E_{\mathbb W}}}R_i
+
\frac{1}{k}
\sum_{\substack{1\leq i\leq k\\ i\notin E\\ i\notin E_{\mathbb W}}}R_i .
\end{equation}
By \eqref{Rk0}, the first term on the right-hand side of
\eqref{criticalregret} is zero.

We now consider the second term. Recall that $
\zeta_i=\min_{1\leq j\leq m}
\left|\phi_i^{\mathrm T}\theta-c_{ji}\right|.$
If $i\notin E_{\mathbb W}$, then $
\zeta_i\leq |\phi_i^{\mathrm T}\tilde{\theta}_i|.$ Thus, for any $\eta>0$,
\begin{equation}\label{critical_bound}
\begin{aligned}
\frac{1}{k}
\sum_{\substack{1\leq i\leq k\\ i\notin E\\ i\notin E_{\mathbb W}}}R_i
&\leq
\frac{m}{k}\sum_{i=1}^k
\operatorname{I}_{\{\zeta_i\leq |\phi_i^{\mathrm T}\tilde{\theta}_i|\}}  \\
&\leq
\frac{m}{k}\sum_{i=1}^k
\operatorname{I}_{\{|\phi_i^{\mathrm T}\tilde{\theta}_i|>\eta\}}
+
\frac{m}{k}\sum_{i=1}^k
\operatorname{I}_{\{\zeta_i\leq \eta\}}  \\
&\leq
\frac{m}{\eta^2 k}\sum_{i=1}^k
(\phi_i^{\mathrm T}\tilde{\theta}_i)^2
+
\frac{m}{k}\sum_{i=1}^k
\operatorname{I}_{\{\zeta_i\leq \eta\}} .
\end{aligned}
\end{equation}
By \eqref{avgphierr}, we have
\begin{equation}
\limsup_{k\to\infty}
\frac{1}{k}
\sum_{\substack{1\leq i\leq k\\ i\notin E\\ i\notin E_{\mathbb W}}}R_i
\leq
m\limsup_{k\to\infty}
\frac{1}{k}\sum_{i=1}^k
\operatorname{I}_{\{\zeta_i\leq \eta\}},
\quad \text{a.s.}
\end{equation}
Letting $\eta\downarrow0$ and using the  condition in Theorem
\ref{regretThm}, 
\begin{equation}\label{final0}
\lim_{k\to\infty}
\frac{1}{k}
\sum_{\substack{1\leq i\leq k\\ i\notin E\\ i\notin E_{\mathbb W}}}R_i
=0,\quad \text{a.s.}
\end{equation}
Combining \eqref{first0}, \eqref{Rk0}, \eqref{criticalregret}, and
\eqref{final0}, we obtain \eqref{regrettheorem}. This completes the proof.
\hfill $\square$
\end{document}